\documentclass[preprint,epsfig,amssymb,prl,comment]{revtex4}
\usepackage{dcolumn}
\usepackage{graphicx,bm}
\usepackage{amsmath}
\begin{document}
\noindent
{\large \bf Comment on ``Entanglement on Demand through Time Reordering''}\\

In a recent Letter~\cite{avron},
Avron {\em et al.}~discuss a time reordering scheme to achieve
efficient ``across-generation'' of entangled photon pairs in a semiconductor quantum dot with a suppressed biexciton binding energy~\cite{Reimer}. They
demonstrate that the scheme can be implemented
using time delay between the generated photons.
In this Comment, we derive an exact expression for the concurrence of the time delayed photons, and show that the predicted values by Avron {\em et al.}~\cite{avron} are
not correct,
which stems from an approximation
used in their approximate
theoretical analysis, namely their Eq.~(16).
We present the
corrected result for time delayed photons, and predict that
the nominal concurrence for generated photons is 0.5 and never above 0.73 (cf.\,0.78 and 1, by Avron {\em et al.}~\cite{avron}), even if one optimally manipulates the biexciton and exciton decay rates. We also discuss a conditional method for achieving higher values of entanglement in this scheme.

The state of the photon pair emitted in biexciton-exciton cascade decay is given by
$
|\psi\rangle=\sum_{k,l}[c_{kl}|1_k\rangle_x|1_l\rangle_x+d_{kl}|1_l\rangle_y|1_k\rangle_y] ,
$
where in each term the first and second kets represent the photon generated in the biexciton and the exciton decays, respectively, and the suffix  labels the polarization. The coefficients $c$ and $d$ are given by
\begin{eqnarray}
c_{kl}\,\,[d_{lk}]=\frac{\sqrt{\Gamma_u\Gamma/2\pi^2}}{(\omega_k+\omega_l-\omega_u+i\Gamma_u)(\omega_l-\omega_{x[y]}+i\Gamma)},
\end{eqnarray}
where for perfect color matching, the biexciton frequency $\omega_u$ is equal to the sum of the exciton frequencies $\omega_x$ and $\omega_y$, and $\Gamma_u$ and $\Gamma$ are the spectral half widths of the biexciton and exciton, respectively. The normalized off-diagonal element of the density matrix is given by
\begin{equation}
\gamma=\int\int c^*_{kl}d_{kl}W_{\rm opt}(\omega_k,\omega_l)\,d\omega_kd\omega_l,
\label{gm}
\end{equation}
where $W_{\rm opt}=\exp\{i\tan^{-1}[(\omega_l-\omega_x)/\Gamma]-i\tan^{-1}[(\omega_k-\omega_y)/\Gamma]\}$ is an additional phase inserted by Avron {\em et al.}~\cite{avron} for optimizing value of $\gamma$. Expanding
the inverse $\tan$ terms, one gets
 $W_{\rm opt}=\exp[i(\omega_l-\omega_x)/\Gamma - i(\omega_k-\omega_y)/\Gamma
 + \cdots]$,
 where the `$\cdots$' labels higher-order terms.
Any practical scheme to achieve additional phase $W_{\rm opt}$, as also highlighted by Avron {\em et al.}~\cite{avron},  would use a
simple
optical delay for the photons that are generated in the biexciton cascade. This
corresponds to precisely the first-order expansion above, but not the full expression (cf.\,the approach used in~\cite{avron})--which
would be
tantamount to an almost impossible delay scheme.
Neglecting constant phases, and introducing an optical delay, $t_0$, we choose $W_{\rm opt}=\exp[-i(\omega_k-\omega_l)t_0]$ in Eq.\,(\ref{gm}), and  find:
\begin{equation}
\gamma=\frac{2\Gamma e^{-2\Gamma t_0}}{\Gamma_u}(1-e^{-\Gamma_u t_0}),
\label{result}
\end{equation}
which implies that $\gamma$ is maximum for $\Gamma t_0=\Gamma\ln(1+\Gamma_u/2\Gamma)/\Gamma_u$. The concurrence for the generated state of photons $|\psi\rangle$ equals to $2|\gamma|$. For $\Gamma_u/\Gamma=2$,  the maximum value of $\gamma$ is only 0.25, which is about 30\% less than the values predicted by Avron {\em et al.}~\cite{avron} (see Fig.\,1(a)). Further, for $\Gamma_u/\Gamma\rightarrow0$, the maximum value of $\gamma=1/e$ thus corresponds to $\Gamma t_0=0.5$.
Similar values have also been reported using a numerical optimization approach~\cite{tejedor}.
\begin{figure}[t]
\centerline{\includegraphics[height=2.1in,width=3in]{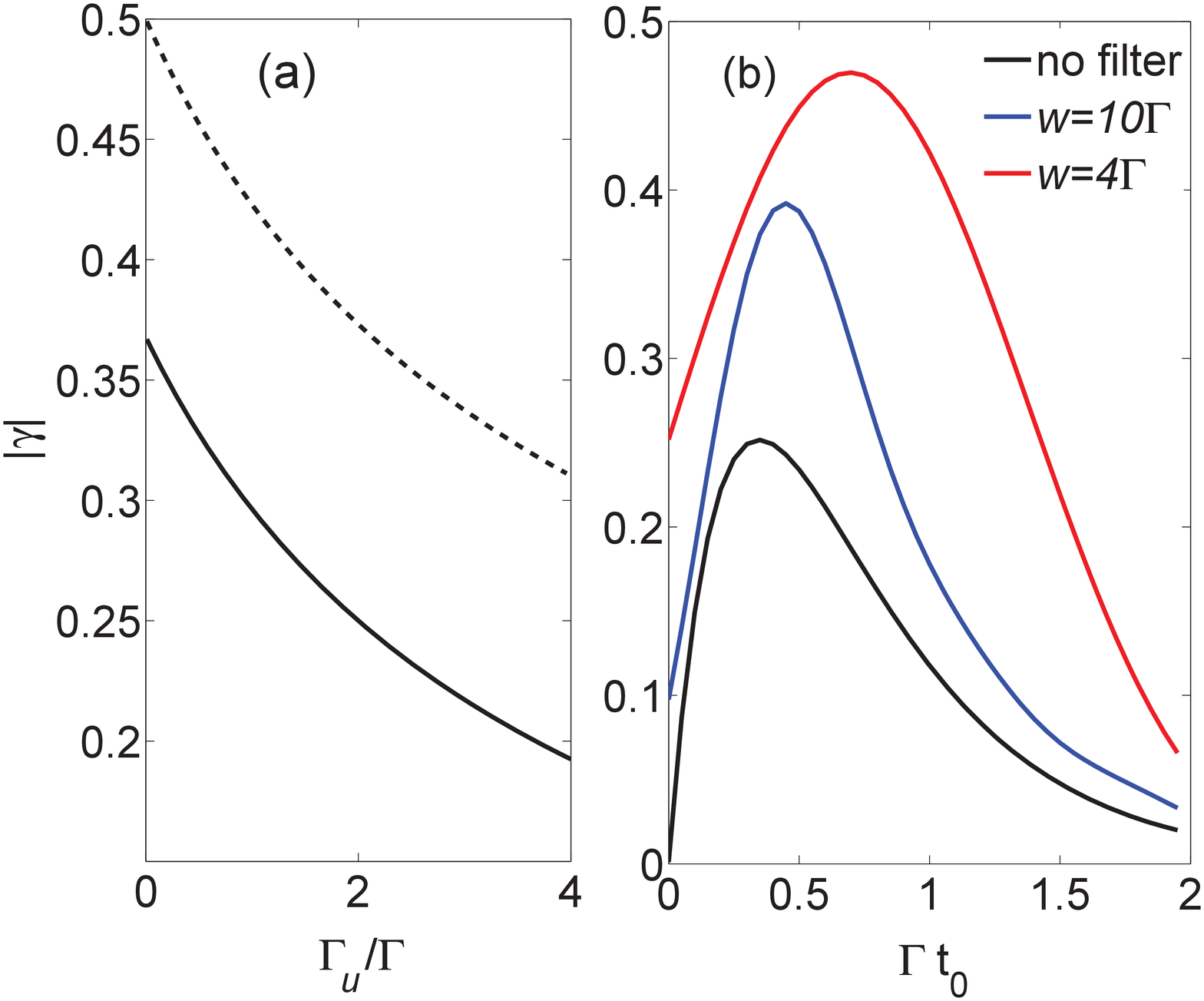}}
\vspace{-0.4cm}
\caption{(color online)
(a) Optimum value of $|\gamma|$ corresponding to time delay $\Gamma t_0=\Gamma\ln(1+\Gamma_u/2\Gamma)/\Gamma_u$ (solid), where the dashed curve shows the optimum values predicted by Avron {\em et al.}~\cite{avron}. (b) The value of $|\gamma|$ for filtered photon pairs, with $\Gamma_u/\Gamma=2$;
the filter function corresponds to a spectral window of width $w=10\Gamma$ (blue), $w=4\Gamma$ (red); the black curve shows results without spectral filter. The conditional
probability
of photon pair generation is $64\%$ (for $w=10\,\Gamma$)  and $33\%$ (for $w=4\,\Gamma$).}
\label{fig1}
\vspace{-0.2cm}
\end{figure}

Finally, we point out that the entanglement can be distilled by using a frequency filter with two spectral windows of width $w$ centered at the frequencies $\omega_x$ and $\omega_y$. The normalized value of $\gamma$ for filtered photons
can be computed by
integrating over the projection operator of the filter~\cite{akopian}.
In Fig.\,\ref{fig1}(b), we confirm that significantly larger values of entanglement in this
time reordering scheme can be achieved by using spectral filter;
however, this
is at the expense
of a reduced probability.
\\
\\
\noindent P. K. Pathak and S. Hughes \\
\indent Department of Physics, Queen's University\\
\indent Kingston, ON K7L 3N6, Canada\\


\vspace{-1.0cm}
\begin{thebibliography}{99}
\vspace{-0.5cm}
\bibitem{avron} J. E. Avron {\em et al.}, Phys. Rev. Lett. {\bf 100}, 120501 (2008)
\bibitem{Reimer} M. E. Reimer {\em et al.}, arXiv:0706.1075v1 (2007).
\bibitem{tejedor}F. Troiani and C. Tejedor, Phys. Rev. B {\bf78}, 155305 (2008).
\bibitem{akopian}N. Akopian {\em et al.},  Phys. Rev. lett {\bf 96}, 130501 (2006).
\end{thebibliography}
\end{document}